\documentclass{article}

    \PassOptionsToPackage{numbers, compress}{natbib}
\usepackage{algorithm}
\usepackage{algpseudocode}
\usepackage{amssymb}
\usepackage{listings}
\usepackage{tabto}
\usepackage[preprint]{neurips_2025}

\usepackage[utf8]{inputenc} 
\usepackage[T1]{fontenc}    
\usepackage{hyperref}       
\usepackage{url}            
\usepackage{booktabs}       
\usepackage{amsfonts}       
\usepackage{amsmath}        
\usepackage{nicefrac}       
\usepackage{microtype}      
\usepackage{graphicx}
\usepackage{wrapfig}

\usepackage{titlesec}
\usepackage{tcolorbox}
\usepackage{tcolorbox}
\title{Friend or Foe}

\author{%
  Oleksandr Cherednichenko\thanks{Integrated Science Lab (IceLab), Department of Mathematics and Mathematical Statistics, Umeå University} \\
  \texttt{oleksandr.cherednichenko@umu.se} \\
  \And
  Josephine Solowiej-Wedderburn\footnotemark[1] \\
  \texttt{josephine.solowiej-wedderburn@umu.se} \\
  \And
  Laura M. Carroll\footnotemark[1]\thanks{Department of Clinical Microbiology, SciLifeLab, Laboratory for Molecular Infection Medicine Sweden (MIMS), Umeå Centre for Microbial Research (UCMR), Umeå University} \\
  \texttt{laura.carroll@umu.se} \\
  \And
  Eric Libby\footnotemark[1] \\
  \texttt{eric.libby@umu.se} \\
}

\begin{document}

\maketitle

\begin{abstract}
A fundamental challenge in microbial ecology is determining whether bacteria compete or cooperate in different environmental conditions.
With recent advances in genome-scale metabolic models, we are now capable of simulating interactions between thousands of pairs of bacteria in thousands of different environmental settings at a scale infeasible experimentally.
These approaches can generate tremendous amounts of data that can be exploited by state-of-the-art machine learning algorithms to uncover the mechanisms driving interactions.
Here, we present Friend or Foe, a compendium of 64 tabular environmental datasets, consisting of more than 26M shared environments for more than 10K pairs of bacteria sampled from two of the largest collections of metabolic models.
The Friend or Foe datasets are curated for a wide range of machine learning tasks---supervised, unsupervised, and generative---to address specific questions underlying bacterial interactions.
We benchmarked a selection of the most recent models for each of these tasks and our results indicate that machine learning can be successful in this application to microbial ecology.
Going beyond, analyses of the Friend or Foe compendium can shed light on the predictability of bacterial interactions and highlight novel research directions into how bacteria infer and navigate their relationships.
\end{abstract}


\section{Introduction}
\label{intro}
A quintessential aspect of ecology is the nature of interactions between living organisms.
For many large (macro-scale) organisms, their interactions are relatively well-defined and fixed, e.g. cats eat mice.
In contrast, bacterial interactions are often context-dependent such that the same pair of organisms can interact differently, e.g. compete or cooperate, depending on the chemical resources present in their environment \cite{Solowiej-Wedderburn2025-ar, qiao_nutrient_2023, martino_oxidative_2024, domeignoz2020microbial, vasse2024killer}. 
Yet, it is unknown to what extent bacteria can actually recognize friend from foe based on the information they can measure.

The challenges bacteria face in distinguishing friend from foe are multiple. 
For one, a primary form of interaction between bacteria is indirect, via manipulation of the environment. 
When bacteria compete, this interaction occurs largely through them consuming the same resource \cite{hibbing2010bacterial,ghoul2016ecology}.
When bacteria cooperate, it often occurs because one species produces a waste product that another one consumes \cite{wintermute2010emergent,harcombe2018evolution,d2018ecology,libby2019syntrophy,hammarlund2019shared}.
Thus, it may not be immediately evident what type of interaction is occurring.
A potentially bigger challenge is that to definitively determine whether interactions are competitive or cooperative, there needs to be a measure of comparison---usually the growth of a species in the absence of the other. 
In this context, competition occurs when the presence of one species causes another to grow more slowly, and cooperation is the converse.
Yet, bacteria do not typically have access to this comparative information.
Moreover, bacteria grow at different rates in different environments, so it is unclear if a species is growing faster or slower due to the presence of another species, or if it is simply because the environment is different.

While there are challenges to determining interactions, bacteria have access to a potentially large collection of data \cite{harapanahalli2015chemical}.
For example, they can sense the concentrations of hundreds of compounds in their environment \cite{elston2023flipping, piepenbreier2017transporters}.
Given evolutionary time scales, bacteria could evolve ways of sensing changes in particularly informative compounds and integrating this information--- via gene regulation \cite{libby2007noisy}--- to infer interactions.
Inferring the type of interaction would be beneficial for species because it could be linked to actions that increase fitness.
For instance, if bacteria infer they are competing with another species, they may then start to make toxins or other weaponry \cite{granato2019evolution}.
Alternatively, it could be that inferring interactions for some species may be too unreliable, leading to the evolution of heuristic strategies. 
Thus, the nature of the inference problem may inform how bacteria sense their environment and respond to other species.

We explore the inference problem by constructing a data set of bacterial interactions with information about specific species and their chemical environments.
We collect this data by using a computational approach featuring genome-scale metabolic models that allows us to significantly expand the quantity of information as compared to traditional wet-lab approaches \cite{Solowiej-Wedderburn2025-ar}.
We call our compendium of data sets \textbf{Friend or Foe} and present it in a tabular format to provide a transparent platform for investigating microbial interactions with existing machine learning methods.
Specifically, Friend or Foe provides data for a variety of machine learning frameworks, including supervised learning, unsupervised learning, transfer learning, and generative modeling. 
For each framework, we selected current state-of-the-art tabular machine learning models and benchmarked them on our datasets. 
\section{Related work}
\label{related}
The question of how a pair, or community, of bacteria will interact is an active area of study.
For some species, the question can be addressed by simply growing them together in a laboratory setting and observing what happens to their populations.
A common outcome of these experiments is that many species compete and few cooperate \cite{palmer_bacterial_2022}; however, conditions in the lab are not representative of species' natural environments.
In addition, the vast majority of bacteria cannot be grown in the lab, presumably because they are engaged in obligate interactions and receive essential resources from other species \cite{pande2017bacterial,kost_metabolic_2023}.
This would suggest that cooperation is common, though such obligate interactions can also arise in parasitic interactions \cite{drew2021microbial}.
Besides observational studies, there is a growing body of research that uses metabolic data to identify the specific molecules driving interactions \cite{klitgord_environments_2010, levy_metabolic_2013, zelezniak_metabolic_2015, machado_polarization_2021}.
This data can be coupled to mathematical models and metabolic models to predict interactions and the population dynamics of bacterial species within communities \cite{hammarlund2019shared}.
Because these approaches require detailed information to inform the modeling, they are typically directed towards socially-relevant communities, e.g. those related to the human gut \cite{heinken_anoxic_2015, magnusdottir_generation_2017}. 

A key tool in predicting microbial interactions has been genome-scale metabolic models.
Recent computational algorithms can take the genome of a bacterium and predict the chemical reactions it can perform as part of its metabolism.
These models can be analyzed by mathematical approaches such as flux balance analysis to predict an organism's growth rate in different chemical environments.
Importantly, metabolic models typically lack information concerning gene regulation or chemical reaction kinetics.
An underlying assumption is that species maximize growth and are constrained by the availability of resources and their repertoire of chemical reactions.
Despite the limitations of metabolic models, their predictions have aligned well with experimental data from species that can be grown in laboratory conditions \cite{harcombe2014metabolic}.
For other species, it is common practice to assume that their models may not be as accurate and to use them more conservatively, e.g. for more general ecology/evolution questions \cite{libby_metabolic_2023,smith2021,libby2019syntrophy,souza2024modeling}.
The Friend or Foe data compendium can be used for such general questions.
\section{Friend or Foe compendium construction}
\label{construction}
\paragraph{Data collection}
We obtained metabolic models from two of the largest collections, AGORA \cite{agora} and CARVEME \cite{carveme}.
One salient difference between the collections is the choice of species; those in AGORA are found in the human gut, while those in CARVEME are sequences in NCBI RefSeq (release 84, \cite{pruitt_ncbi_2007}), which come from a range of environments, including aquatic and terrestrial environments.
We followed the procedure outlined in \cite{el1} to format the metabolic models. 
For each model, there is a unique stoichiometric matrix in which rows correspond to chemical compounds and columns represent metabolic reactions. 
\paragraph{Generating the environments}
To generate a large collection of environments with different types of bacterial interactions, we follow the procedure described in \cite{Solowiej-Wedderburn2025-ar}.
Here, we briefly summarize the main steps to provide context for the datasets in our compendium.
First, we determine the growth rate $\lambda$ of bacteria by using the established technique of flux balance analysis.
This approach formulates metabolic growth as a linear program in which the rate that each reaction is used (the ``fluxes'') are computed so they maximize the amount of biomass produced by the organism (see Algorithm \ref{al1}).
Fluxes are constrained by upper and lower bounds to indicate directionality in reactions and prevent infinite growth.
Fluxes also must satisfy constraints that determine which compounds must be balanced (i.e., no net change in their amounts) and which can be unbalanced (those exchanged with the environment).
\par
\begin{minipage}{0.48\textwidth}

    \begin{algorithm}[H]
    \caption{Flux Balance Analysis for a pair}
    \label{al1}
    \begin{algorithmic}[1]
        \State \textbf{Require} S-matrices for microbes $M_i$ and $M_j$: $S_{M_i} =[S_{E_i}, S_{I_i}]$; $S_{M_j}=[S_{E_i}, S_{I_j}]$
        \State \textbf{Require} Bounds on fluxes and compounds: $\mathbf{l}_{v}, \mathbf{u}_{v}, \mathbf{l}_{\Delta}, \mathbf{u}_{\Delta}$;
        \State \textbf{Define} S-matrix and flux vectors for a pair \begin{equation*}
    S_{M_i + M_j} = 
    \begin{bmatrix}
        S_{E_i} & S_{E_j} \\
        S_{I_i} & 0 \\
        0 & S_{I_j}
    \end{bmatrix}
\end{equation*}
\begin{equation*}
\mathbf{v}^* = \begin{bmatrix} v_{M_i} ; v_{M_j} \end{bmatrix}^{T}
\end{equation*}
\begin{equation*}
\mathbf{\Delta}^* = \begin{bmatrix} \Delta_{E_i} + \Delta_{E_j} ; \Delta_{I_i} ; \Delta_{I_j} \end{bmatrix}^{T}
\end{equation*}
\vspace{0.005mm}
\State \textbf{Solve} for $M_i$ with $M_j$ no worse than alone

\begin{equation*}
\begin{aligned}[c]
\max \left(\lambda^{*}_{M_{i,j}}\right)\\
S_{M_{i}+M_{j}} \cdot \mathbf{v}^* = \mathbf{\Delta}^*\\
\mathbf{l}_{v} \leq \mathbf{v}^* \leq \mathbf{u}_{v}\\
\mathbf{l}_{\Delta} \leq \mathbf{\Delta}^* \leq \mathbf{u}_{\Delta}\\
\lambda_{M_{j,i}} \geq \lambda^{\dagger}_{M_{j}} - \epsilon
\end{aligned}
\qquad
\begin{aligned}[c]
\max \left(\lambda^{\dagger}_{M_{j}}\right)\\
S_{M_{j}} \cdot \mathbf{v}^\dagger = \mathbf{\Delta}^\dagger\\
\mathbf{l}_{v} \leq \mathbf{v}^\dagger \leq \mathbf{u}_{v}\\
\mathbf{l}_{\Delta} \leq \mathbf{\Delta}^\dagger \leq \mathbf{u}_{\Delta}\\
\quad
\end{aligned}
\end{equation*}
        \State \textbf{return} $\mathbf{v}^*, \mathbf{\Delta}^*, \lambda^*_{M_i}, \lambda^*_{M_j}, \lambda^*_{M_{j,i}}, \lambda^*_{M_{i,j}}$
    \end{algorithmic}
    \end{algorithm}
\end{minipage}%
\hfill
\begin{minipage}{0.48\textwidth}
    \begin{algorithm}[H]
    \caption{Generating the environments}
    \label{al2}
    \begin{algorithmic}[1]
\State \textbf{Require} number of pairs $N_p$, 
concentration $c$, pair of microbes ($S_{M_i}$, $S_{M_j}$), compound set $C$
\For{$p \leftarrow 1$ to $N_p$} \\
    \color{blue}
    \# Generate random environments 
    \color{black}
    \State Find usable compounds $C_U$
    \State Find essential compounds $C_E$
    \State Sample $q \sim \mathcal{U}[C_U \setminus C_E]$; set $\mathbf{l}_{\Delta}(q) = c$
    \State Execute Algorithm \ref{al1} for $p$
    \\
    \color{blue}
    \# Identify interaction 
    \color{black}
    \State Compare $\lambda^*_{M_i}, \lambda^*_{M_j},
    \lambda^*_{M_{j,i}}, \lambda^*_{M_{i,j}}$\footnotemark; 
    \State Construct environment $x \in \mathbb{R}^{1\times C}$

    \State Construct target $y \in \mathbb{R}$
    \State Update interaction summary

    \If{desired interaction (Table~\ref{tab:abbreviations})}
        \State Update counter $n$
        \State Update $x$ with $\mathbf{l}_{\Delta}$ and $y$
        \State Fuse $x$ into $X \in \mathbb{R}^{n \times C}$
    \EndIf

    \State \textbf{When} desired number of interactions or 
    
    upper search threshold reached
        \State \textbf{break}

\EndFor
\State \Return $\mathcal{D}(X, y)$ or $\mathcal{D}(X)$
    \end{algorithmic}
    \end{algorithm}
\end{minipage}
\footnotetext{We compare the growth rates of each microbe alone and in co-culture to identify interaction types. See Supplementary D for details.}
\par
These constraints are represented by the product of the stoichiometric matrix $S_M=[S_{E}, S_{I}]$ and a vector of fluxes $\mathbf{v}$, which gives a right-hand-side vector $\mathbf{\Delta}$ of changes in compound concentrations.
For faster computations, we separated the stoichiometric matrices into internal $S_I$ and extracellular $S_E$ compartments.
It is assumed that compounds in the internal compartment (inside the cell) must be balanced so their upper $\mathbf{u_\Delta}$ and lower bounds $\mathbf{l_\Delta}$ are zero, while extracellular compounds can be unbalanced so at least one of their bounds will be non-zero.
Specifically, negative bounds correspond to compounds that can be imported from the environment into the cell, while positive bounds are the converse.
Hence, we refer to compounds with negative lower bounds $\mathbf{l_\Delta}$ in the extracellular compartment as the `environment’.
\begin{table}[ht]
    \caption{Datasets for classifying interactions. \texttt{BC} and  \texttt{MC} stand for Binary Classification and Multiclassification, respectively. The notation (\(+\times\)) means that bacteria \(+\) can grow alone in the environment, while  \(\times\) cannot. Notation (\(\times\times\)) means that neither can survive alone.}
    \centering
    \small
    \begin{tabular*}{\linewidth}{l @{\extracolsep{\fill}} c @{\extracolsep{\fill}} l}
        \toprule
        \multicolumn{1}{c}{\textnormal{Notation}} & \multicolumn{1}{c}{\textnormal{Number of Classes $k$}} & \multicolumn{1}{c}{\textnormal{Description}} \\
        \midrule
        \textnormal{BC-I}  & 2 & \textnormal{Facultative Cooperation \textbf{vs} Competition}\\
        \textnormal{BC-II} & 2 & \textnormal{Obligate (\(+\times\)) Cooperation \textbf{vs} Facultative Cooperation} \\
        \textnormal{BC-III}   & 2 &  \textnormal{Obligate (\(+\times\)) Cooperation \textbf{vs} Competition} \\
        \textnormal{BC-IV}  & 2 & \textnormal{Obligate (\(\times+\)) Cooperation \textbf{vs} Facultative Cooperation} \\
        \textnormal{BC-V} & 2 & \textnormal{Obligate (\(\times+\)) Cooperation \textbf{vs} Competition} \\
        \textnormal{MC-I} & 3 & \textnormal{Obligate (\(\times\times\)) \textbf{vs} Facultative \textbf{vs} Competition} \\
        \textnormal{MC-II}  & 3 & \textnormal{Obligate (\(+\times\))  \textbf{vs} Obligate (\(\times+\)) \textbf{vs} Obligate (\(\times\times\))}\\
        \textnormal{MC-III} & 5 & \textnormal{All classes present} \\
        \bottomrule
    \end{tabular*}
    \label{tab:abbreviations}
\end{table}
Since a driving question for our analyses was the role of the environment in determining bacterial interactions, we needed a large collection of environments and interactions. 
We generated different environments by changing the lower bounds of the right-hand-side vector following Algorithm \ref{al2}.
By comparing the growth rates of bacteria alone and in pairs using the approach in \cite{Solowiej-Wedderburn2025-ar}, we identified those interactions described in Table \ref{tab:abbreviations}.

If there were no way for both species to grow at least as fast as they did alone, then there is competition.
Alternatively, if there is a way for both species to grow faster, then it is cooperation. 
Other logical possibilities, e.g. neutral interactions, were disregarded for this analysis.
To avoid biasing the search of environments for different interactions, we randomly sampled compounds from a uniform distribution; however, this had computational drawbacks, as the majority of environments sampled resulted in no growth (see Friend or Foe pipeline and Supplementary D).
\paragraph{Essential and additional compounds}
When generating environments for a pair of bacteria, we first identified essential compounds $C_E$, which must always be present for the pair to grow.  
While necessary, essential compounds alone are not sufficient for growth.
We thus identified those extracellular compounds that could be used by at least one of the pair $C_U$ and sampled a fixed  number (50 or 100) of these usable compounds.

\begin{figure}[ht]
	\centering 
 \includegraphics[width=\textwidth]{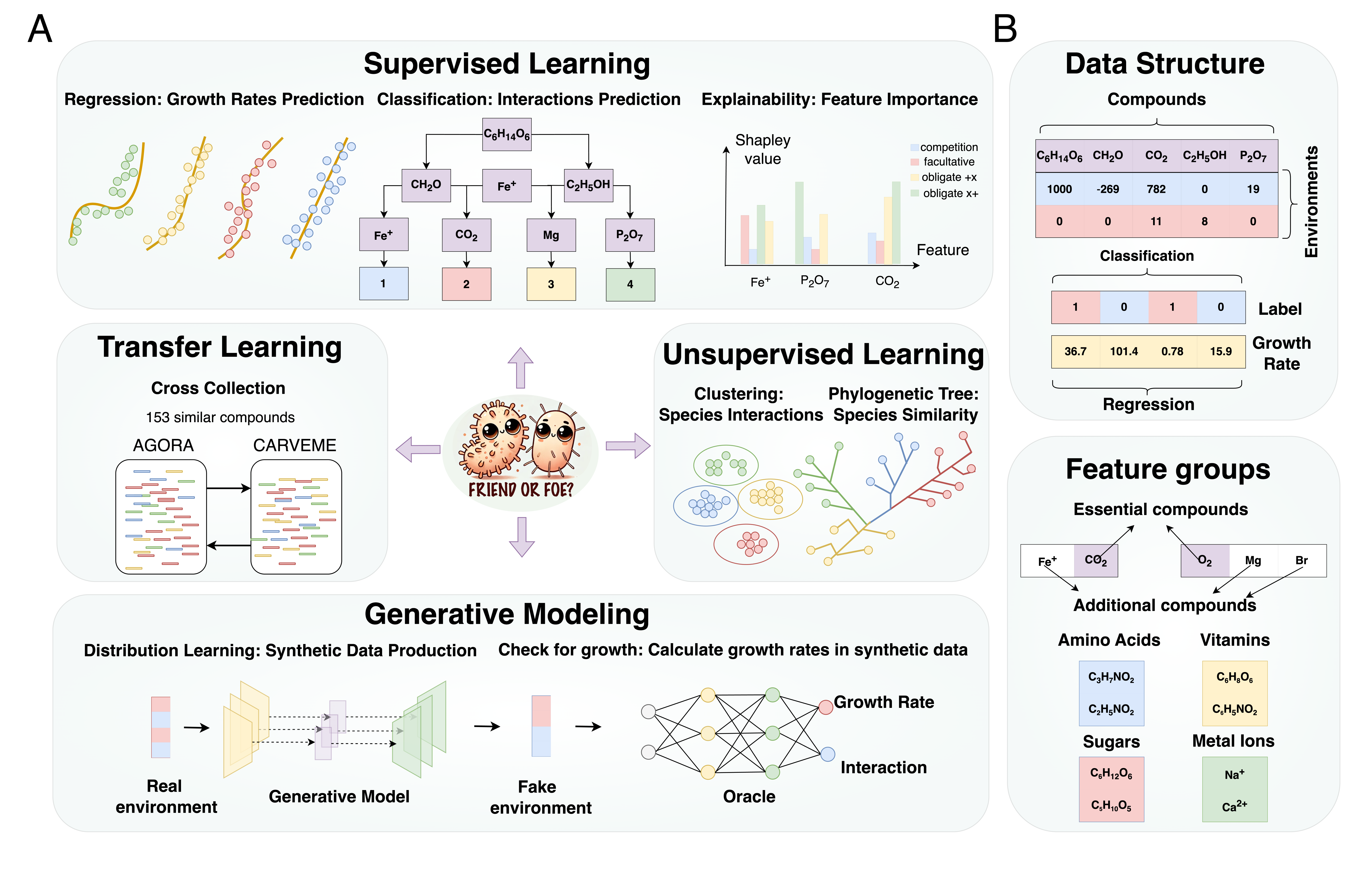}

	\caption{\textbf{A. Friend or Foe,} 
    a universal compilation of datasets describing interactions in bacteria.
    Schematic outlines four possible machine learning tasks that could be used to probe different eco-evolutionary questions. 
    \\
    \textbf{B. Datasets are structured in a tabular domain.} Each dataset is a table with columns corresponding to names of compounds and reactions present in a particular collection. Essential compounds are such compounds without which  a pair cannot grow together. Additional compounds are added to the environment to ensure richness and diversity. 
    }

	\label{main}%
\end{figure}

\paragraph{Types of microbial interactions}
Algorithm \ref{al2} returns datasets $\mathcal{D}(X,y)$ or $\mathcal{D}(X)$, which contain a matrix of environments $X$ with or without targets $y$.
These targets could either be discrete data for the interaction in each environment, or continuous values for the growth rates.
\begin{wraptable}{r}{0.4\textwidth}
    \centering
    \small
    \begin{tabular}{l c}
        \toprule
        \textbf{Chemical class} & \textbf{Counts} \\
        \midrule
         Amino Acids  & 69 \\
        Sugars & 56 \\
         Vitamins & 54 \\
        Carboxylic Acids & 51 \\
         Lipids \& Fatty Acids & 47 \\
          Metal Ions & 22 \\
        Other & 125 \\
        \bottomrule
    \end{tabular}
    \caption{A classification of  compounds based on chemical structure.}
    \label{tab:metabolites}
\end{wraptable}
In this work, we consider five different types of interactions (i.e., classes).
The first is \texttt{Competition} in which species negatively affect each other by competing for limited compounds so that the growth rate of at least one species is worse compared to when it grows alone.
The remaining four interactions are types of cooperation in which both bacteria have higher growth rates when grown together versus alone.
In \texttt{Facultative cooperation}, the relationship is not essential such that each species can also grow independently. 
In \texttt{Obligate} (\(+\times\)) or (\(\times+\)) \texttt{cooperation}, the \(\times\) species cannot survive without the \(+\) partner.
Finally, in \texttt{Obligate} (\(\times\times\)) \texttt{cooperation}, neither species can survive independently.
\par
\paragraph{Tabular structure}
Datasets (constructed with Algorithm \ref{al2}) are curated in a tabular domain, where columns represent compounds and rows represent environments.
Compounds (features) are additionally categorized by their chemical similarity (Table \ref{tab:metabolites}). 

\par
\paragraph{Friend or Foe pipeline}
The curated datasets make it possible to address a diverse range of questions concerning bacterial interactions.
For illustration, we have selected four machine learning frameworks to probe different questions (Figure \ref{main}).
The main motivation for this compendium concerns whether a species of bacteria can infer its interactions using only the information of which compounds are present in the environment.
This question can be addressed with a supervised learning framework using labeled datasets $\mathcal{D}(X,y)$, generated by running Algorithm \ref{al2}. 
We organized the data into different datasets depending on whether the classification is binary \texttt{BC} or multiclassification \texttt{MC} (Tables \ref{tab:abbreviations}, \ref{extended_table_1}). 
\par
\begin{wrapfigure}{r}{0.4\textwidth} 
    \centering
    \includegraphics[width=0.8\linewidth]{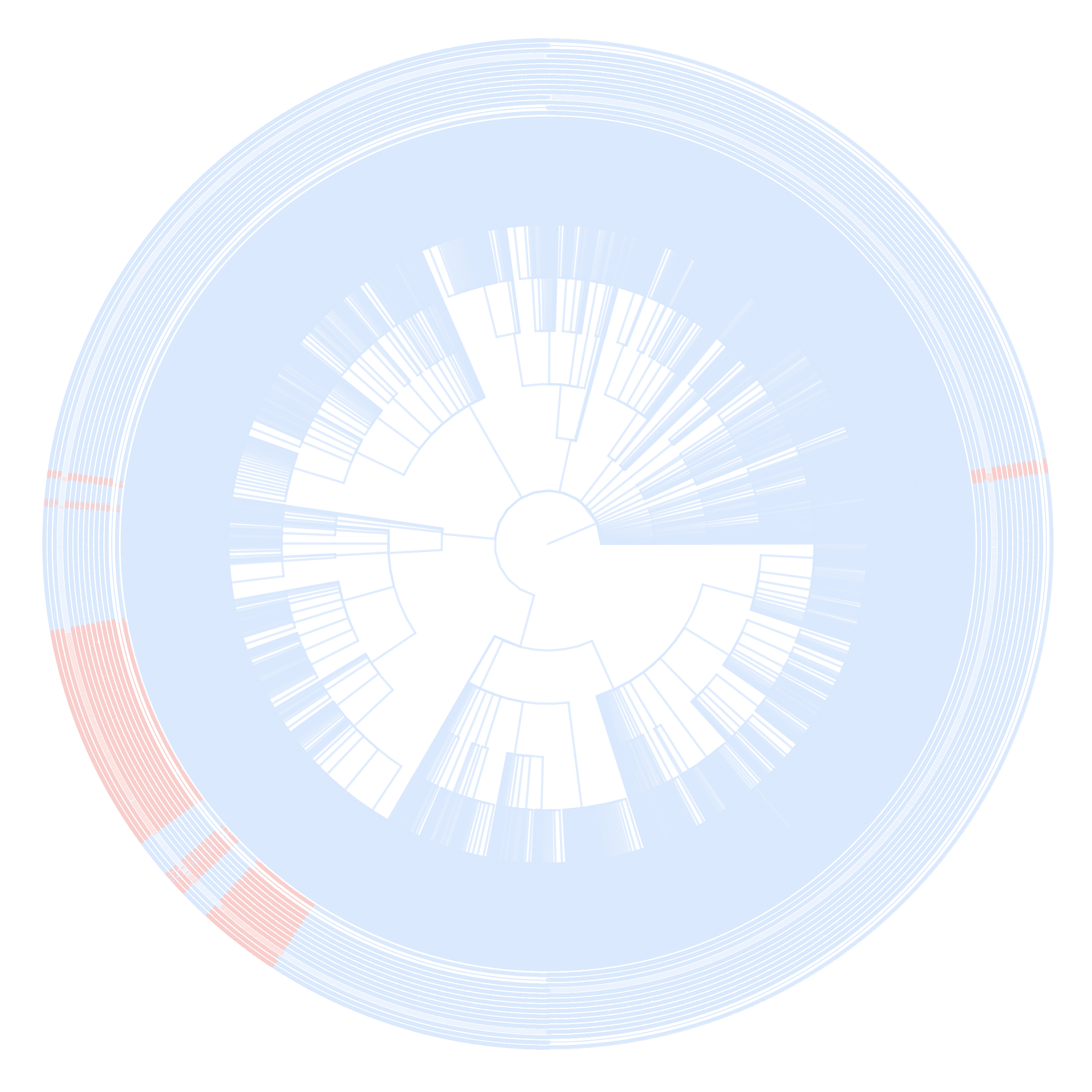}
    \caption{The taxonomic tree for the CARVEME species collection, where blue nodes represent CARVEME-specific species, and red nodes indicate species shared with AGORA.
    }
    \label{tree}
\end{wrapfigure}
\par

\begin{figure}[ht]
	\centering 
 \includegraphics[width=\textwidth]{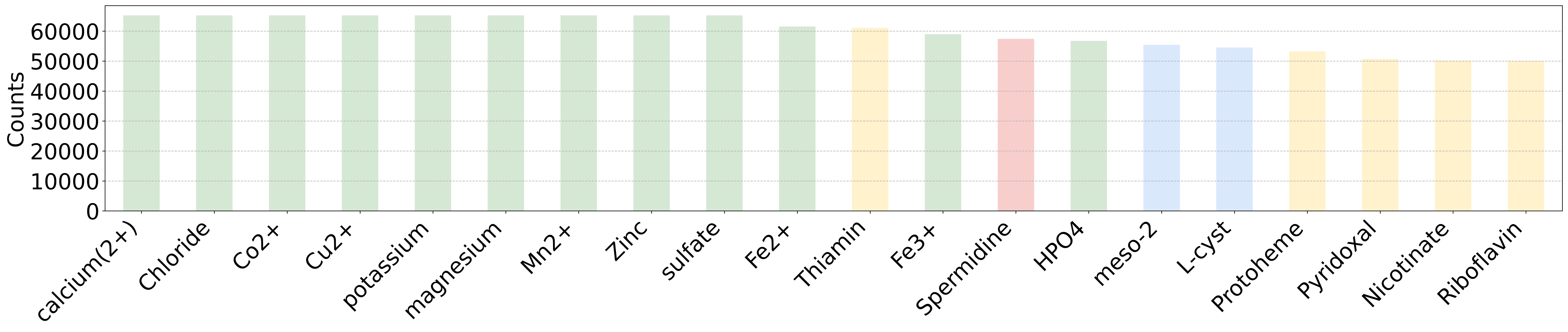}

	\caption{The barplot shows the top 20 chemical compounds that occur the most in both collections. Colors show the specific chemical class (Table \ref{tab:metabolites} and Figure \ref{main}).}

	\label{j}%
\end{figure}

Besides predicting interactions between bacteria, another question we can address is whether it is possible to predict bacterial growth rates from environmental matrix $X$.
One approach to this question uses a supervised learning framework with interaction-specific datasets labeled with growth rate outputs from Algorithm \ref{al2} (see \texttt{GR-I}, \texttt{GR-II} and \texttt{GR-III}, Table \ref{extended_table_1}).
We could then use a supervised learning framework to quantify the feature importance of different compounds in the environment, using either the specific compound names (see, e.g., Figure \ref{j}) or their classifications (as in Table \ref{tab:metabolites}).
If these approaches are successful, it would imply that machine learning algorithms can uncover the main drivers (feature importance) determining bacterial growth in different environments, and bacterial species may rely on these compounds as inputs to their decision making.
If unsuccessful, it would suggest that the inference problem is too complex to solve in general and species might rely on heuristic approaches to their interactions and growth decisions.
The previous questions made use of supervised learning, but there are other questions that make use of our dataset, which may be suitable for other types of learning approaches.
For example, we may want to know the extent to which species similarity is informative. 
We constructed a taxonomic tree to organize the evolutionary relationships between species (Figure \ref{tree}).
Using this data, we can test the hypothesis that closely related organisms may compete more often for resources because they share common metabolisms.
Conversely, pairs of distant organisms may have more scope for cooperation.
We can integrate this task into an unsupervised clustering problem.
To test our hypothesis, we created datasets \texttt{US-I} and \texttt{US-II} (Table \ref{extended_table_1}) without ground-truth labels of the interactions. 
We can then employ clustering algorithms to see if the environments are separable into groups that could represent a particular interaction between species, taking into account taxonomy.

\begin{table}[ht]
\caption{The compendium consists of 64 datasets. The \texttt{Samples} represents the amount of environments for each \texttt{Task} with additional 100/50 compounds. The \texttt{Model} is the name of model that shows the best performance. Data was split in train/val/test with a ratio 6/2/2. \texttt{Group} indicates the number of additional compounds. \texttt{Collection} refers to AGORA \cite{agora} or CARVEME \cite{carveme}. AGORA datasets have 424 features (unique chemical compounds), while CARVEME datasets have 499.}
\label{extended_table_1}
\vskip 0.15in
\begin{center}
\begin{small}
\begin{sc}
\begin{tabular*}{\textwidth}{l@{\extracolsep{\fill}} ccc ccc}
\toprule
\textnormal{Task} & \textnormal{Samples} & \textnormal{Model} & \textnormal{Group} & \textnormal{Metric} & \textnormal{Collection} \\
\midrule
\textnormal{BC-I}   & 326\,331 \,/\ 89\,500 & \textnormal{TabM} & 100 \,/\ 50 & \textnormal{Acc}  \,/\ \textnormal{MCC} & \textnormal{AGORA} \\
\textnormal{BC-II}  & 624\,743 \,/\ 547\,175 & \textnormal{FT-Transformer} & 100 \,/\ 50 & \textnormal{Acc} \,/\ \textnormal{MCC} & \textnormal{AGORA} \\
\textnormal{BC-III} & 509\,298 \,/\ 491\,167 & \textnormal{TabM} & 100 \,/\ 50 & \textnormal{Acc} \,/\ \textnormal{MCC} & \textnormal{AGORA} \\
\textnormal{BC-IV}  & 339\,006 \,/\ 198\,485 & \textnormal{TabM} & 100 \,/\ 50 & \textnormal{Acc} \,/\ \textnormal{MCC} & \textnormal{AGORA} \\
\textnormal{BC-V}  & 223\,561 \,/\ 142\,477 & \textnormal{TabM} & 100 \,/\ 50 & \textnormal{Acc} \,/\ \textnormal{MCC} & \textnormal{AGORA} \\
\textnormal{MC-I} & 730\,186 \,/\ 563\,922 & \textnormal{TabM} & 100 \,/\ 50 & \textnormal{Acc} \,/\ \textnormal{MCC} & \textnormal{AGORA} \\
\textnormal{MC-II}  & 925\,828 \,/\ 1\,074\,570 & \textnormal{TabM} & 100 \,/\ 50 & \textnormal{Acc} \,/\ \textnormal{MCC} & \textnormal{AGORA} \\
\textnormal{MC-III} & 1\,252\,159 \,/\ 1\,164\,072 & \textnormal{TabM} & 100 \,/\ 50 & \textnormal{Acc} \,/\ \textnormal{MCC} & \textnormal{AGORA} \\
\textnormal{GR-I}  & 105\,443 \,/\ 16\,747 & \textnormal{FT-Transformer} & 100 \,/\ 50 & \textnormal{RMSE} \,/\ $r^2$ & \textnormal{AGORA} \\
\textnormal{GR-II} & 220\,888 \,/\ 72\,755 & \textnormal{TabM} & 100 \,/\ 50 & \textnormal{RMSE} \,/\ $r^2$ & \textnormal{AGORA} \\
\textnormal{GR-III} & 403\,855 \,/\ 474\,420 & \textnormal{TabNet} & 100 \,/\ 50 & \textnormal{RMSE} \,/\ $r^2$ & \textnormal{AGORA} \\
\textnormal{TL-I} & 659\,269 \,/\ 257\,420 & \textnormal{TabM} & 100 \,/\ 50 & \textnormal{Acc} \,/\ \textnormal{MCC} & \textnormal{AGORA} \\
\textnormal{TL-II} & 1\,997\,248 \,/\ 1\,580\,917 & \textnormal{TabM} & 100 \,/\ 50 & 
\textnormal{Acc} \,/\ \textnormal{MCC} & \textnormal{AGORA} \\
\textnormal{US-I}  & 12\,002 \,/\ 12\,000 & DBSCAN & 100 \,/\ 50 & \textnormal{DCSI} \,/\ \textnormal{SC} & \textnormal{AGORA} \\
\textnormal{US-II} & 12\,001 \,/\ 3\,000 & DBSCAN & 100 \,/\ 50 & \textnormal{DCSI} \,/\ \textnormal{SC} & \textnormal{AGORA} \\
\textnormal{GEN} & 40\,000 \,/\ 20\,000 & \textnormal{TVAE} & 100 \,/\ 50 & $\alpha$-\textnormal{Precision} \,/\ $\beta$-\textnormal{Recall} & \textnormal{AGORA} \\
\midrule
\textnormal{BC-I}   & 332\,938 \,/\ 167\,918 & \textnormal{TabM} & 100 \,/\ 50 & \textnormal{Acc} \,/\ \textnormal{MCC} & \textnormal{CARVEME} \\
\textnormal{BC-II}  & 399\,751 \,/\ 466\,814 & \textnormal{FT-Transformer} & 100 \,/\ 50 & \textnormal{Acc} \,/\ \textnormal{MCC} & \textnormal{CARVEME} \\
\textnormal{BC-III} & 478\,057 \,/\ 486\,646 & \textnormal{TabM} & 100 \,/\ 50 & \textnormal{Acc} \,/\ \textnormal{MCC} & \textnormal{CARVEME} \\
\textnormal{BC-IV}  & 237\,535 \,/\ 211\,179 & \textnormal{TabM} & 100 \,/\ 50 & \textnormal{Acc} \,/\ \textnormal{MCC} & \textnormal{CARVEME} \\
\textnormal{BC-V}  & 315\,841 \,/\ 231\,011 & \textnormal{TabM} & 100 \,/\ 50 & \textnormal{Acc} \,/\ \textnormal{MCC} & \textnormal{CARVEME} \\
\textnormal{MC-I} & 605\,373 \,/\ 560\,689 & \textnormal{TabM} & 100 \,/\ 50 & \textnormal{Acc} \,/\ \textnormal{MCC} & \textnormal{CARVEME} \\
\textnormal{MC-II}  & 655\,089 \,/\ 922\,678 & \textnormal{TabM} & 100 \,/\ 50 & \textnormal{Acc} \,/\ \textnormal{MCC} & \textnormal{CARVEME} \\
\textnormal{MC-III} & 988\,027 \,/\ 1\,090\,596 & \textnormal{TabM} & 100 \,/\ 50 & \textnormal{Acc} \,/\ \textnormal{MCC} & \textnormal{CARVEME} \\
\textnormal{GR-I}  & 339\,004 \,/\ 93\,875 & \textnormal{FT-Transformer} & 100 \,/\ 50 & \textnormal{RMSE} \,/\ $r^2$ & \textnormal{CARVEME} \\
\textnormal{GR-II} & 205\,622 \,/\ 93\,875 & \textnormal{TabM} & 100 \,/\ 50 & \textnormal{RMSE} \,/\ $r^2$ & \textnormal{CARVEME} \\
\textnormal{GR-III} & 272\,435 \,/\ 392\,771 & \textnormal{TabNet} & 100 \,/\ 50 & \textnormal{RMSE} \,/\ $r^2$ & \textnormal{CARVEME} \\
\textnormal{TL-I} & 494\,249 \,/\ 454\,572 & \textnormal{TabM} & 100 \,/\ 50 & \textnormal{Acc} \,/\ \textnormal{MCC} & \textnormal{CARVEME} \\
\textnormal{TL-II} & 1\,848\,506 \,/\ 1\,729\,659 & \textnormal{TabM} & 100 \,/\ 50 & 
\textnormal{Acc} \,/\ \textnormal{MCC} & \textnormal{CARVEME} \\
\textnormal{US-I}  & 9\,004 \,/\ 8\,000 & DBSCAN & 100 \,/\ 50 & \textnormal{DCSI} \,/\ \textnormal{SC} & \textnormal{CARVEME} \\
\textnormal{US-II} & 11\,000 \,/\ 8\,000 & DBSCAN & 100 \,/\ 50 & \textnormal{DCSI} \,/\ \textnormal{SC} & \textnormal{CARVEME} \\
\textnormal{GEN} & 40\,000 \,/\ 20\,000 & \textnormal{TabDiff} & 100 \,/\ 50 & $\alpha$-\textnormal{Precision} \,/\ $\beta$-\textnormal{Recall} & \textnormal{CARVEME} \\
\bottomrule
\end{tabular*}
\end{sc}
\end{small}
\end{center}
\vskip -0.1in
\end{table}

\par
One challenge that we faced in constructing our Friend or Foe compendium was the rate of failure in generating environments with particular types of interactions. 
For example, for each pair of species, we sampled an average of 414 random environments to find 100 that were viable per pair.
Furthermore, of these 100 viable environments, we only found an average of 9 competitive environments per pair.
It would improve our efficiency if we could modify our random sampling approach to lower the failure rate.
Such a task would be suitable for generative modeling approaches that aim to learn the distribution of a given dataset and use this knowledge to produce synthetic data points.
We created a dataset \texttt{Gen} to evaluate these approaches.
\par
The metabolic models for our datasets come from two distinct collections: AGORA and CARVEME.
So far, all our approaches have been collection-specific, and we have not combined the data from each collection, as they differ in both construction and features.
However, we did identify some overlap: both collections share 153 compounds and 467 organisms (see Supplementary B for details).
To increase the generalization of our results, we used this overlapping information to construct datasets for Transfer Learning \texttt{TL-I} and \texttt{TL-II}.
We performed Transfer Learning for supervised classification and regression, training on the training set from one collection and testing it on the testing set from the other collection.
Fused together, we show a dataset structure of Friend or Foe in Table \ref{extended_table_1}.

\section{Experimental Evaluation}
\label{headings}

\subsection{Benchmarks}
\paragraph{Supervised learning and Transfer learning} 

\begin{figure*}
	\centering 
 \includegraphics[width=\textwidth]{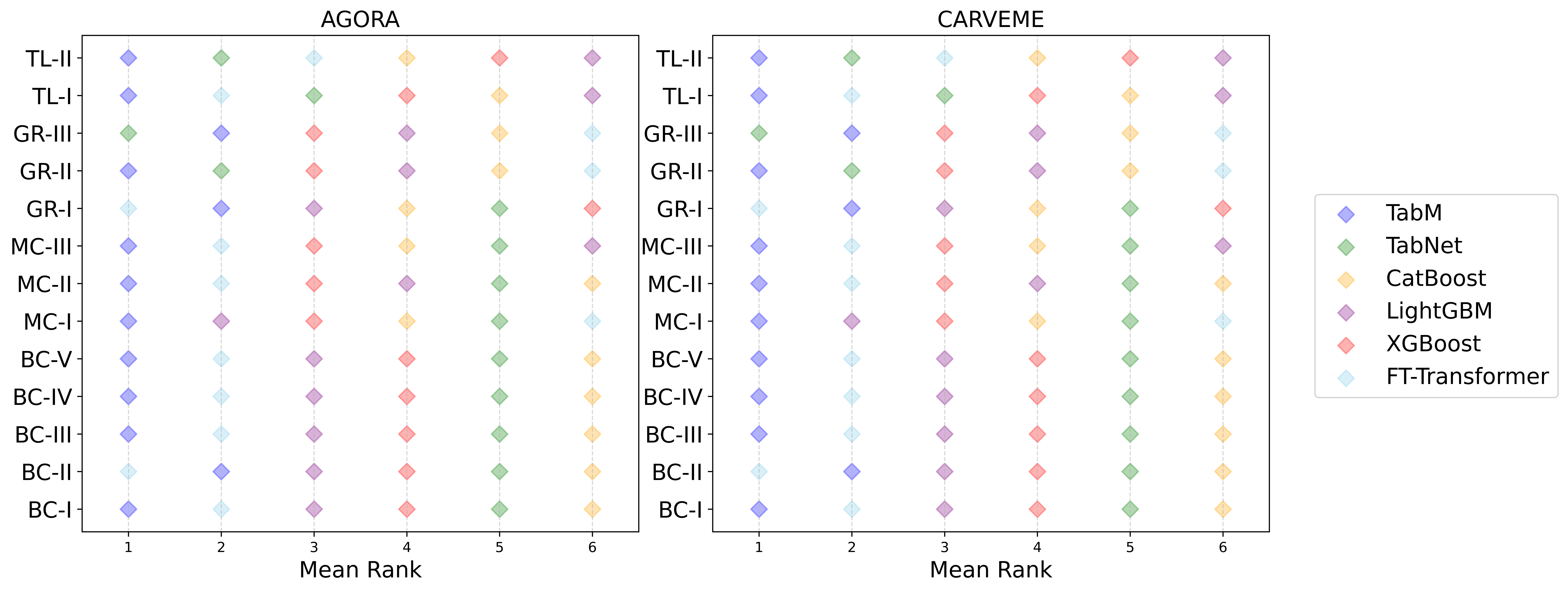}

	\caption{Benchmark of state-of-the-art tabular ML models on Supervised learning and Transfer learning tasks across all datasets from Table \ref{tab:metabolites}.}

	\label{sl}%
\end{figure*}

 We tested different classical machine learning algorithms for our supervised learning approach to address the question of whether bacteria can classify interactions as competition or cooperation.
Namely, we used two different types of algorithms: Gradient Boosting based Decision Trees---XGBoost \cite{xgb}, LightGBM \cite{lgbm} and CatBoost \cite{catboost}---and recent advanced deep learning algorithms---FT-Transformer \cite{gorishniy2022on}, TabM \cite{gorishniy2025tabm}, TabNet \cite{arik2020tabnet}.
For classification tasks (interaction prediction), we used accuracy (Acc) and Matthew correlation coefficient (MCC) to evaluate the algorithms; for regression tasks (growth rate prediction), we used Root mean squared error (RMSE) and determination coefficient $r^2$ as evaluation metrics.
These tasks cover 48/64 datasets in Friend or Foe, as outlined in Table \ref{extended_table_1}. 
In Figure \ref{sl} we plot the rankings of each algorithm based on their performance per dataset. 
The results show that across all classification tasks, the highest ranked algorithm was always a deep learning algorithm, of these TabM often came out best.
Table \ref{sltab} displays the mean MCC and Accuracy metrics averaged across all the corresponding tasks.
These values indicate that all algorithms achieve meaningful predictive performance, successfully capturing patterns in bacterial interactions and growth rates.
We also obtained positive results for the transfer learning problem, addressing the ability to make interaction predictions across the two collections AGORA and CARVEME.
Our results suggest that machine learning algorithms can provide insight into the environmental drivers of bacterial interactions and how bacteria themselves may use this information for interaction strategies.

\begin{table}[ht]
    \caption{Benchmark of supervised models. We evaluated tabular ML models taking on Interaction Prediction (IP), which is a classification task. IP is aggregated classification datasets from Table \ref{extended_table_1}. We measure MCC and Accuracy(Acc) and report mean with std for $15$ runs.}
    \centering
    \small
    \begin{tabular}{lcccccc}
   \\
   \toprule
        & TabM & FT & TabNet & CatB & XGB & LGBM \\
        \midrule
        MCC  $\uparrow$  &  $0.64 \pm 0.03$ &   $0.57 \pm 0.02$ & $0.55 \pm 0.02$ & $0.52 \pm 0.03$ & $0.47 \pm 0.00$ & $0.51 \pm 0.00$ \\
        \midrule
        Acc  $\uparrow$ &  $0.84 \pm 0.02$ & $0.78 \pm 0.06$ & $0.72 \pm 0.01$ & $0.71 \pm 0.03$ & $0.69 \pm 0.01$ & $0.70 \pm 0.02$ \\
        \bottomrule
    \end{tabular}
    \label{sltab}
\end{table}

\paragraph{Unsupervised learning}
We tested four different unsupervised learning algorithms suitable to test our hypothesis that the phylogenetic relatedness of species predicts their likelihood to compete, using datasets \texttt{US-I} and \texttt{US-II}.
We chose two classical methods—-$K$-means and DBSCAN \cite{10.5555/3001460.3001507}—-and two more recent fairness-based methods—FairSC \cite{kleindessner2019guaranteesspectralclusteringfairness} and FairDen \cite{krieger2025fairden}.
In Table \ref{ustab}, we evaluate the algorithms using standard clustering metrics: Density Cluster Separability Index (DCSI) \cite{gauss2025dcsiimprovedmeasure} and Silhouette Coefficient (SC).
Given the relatively small number of samples in these datasets, we only consider these tests as a proof-of-concept for our methodology and intend to expand them in our future pipeline.
A major challenge we faced here was the high proportion of unviable environments (where bacteria could not grow) when randomly sampling.
We anticipate that our generative modeling approach may help this process.

\begin{table}[ht]  
    \caption{Benchmark of unsupervised clustering algorithms. We used DCSI \cite{gauss2025dcsiimprovedmeasure} and SC as clustering metrics. We report mean with std for $10$ runs. }
    \centering
    \small
    \begin{tabular}{lcccc}
        \toprule
        & \multicolumn{2}{c}{\textbf{US-I}} & \multicolumn{2}{c}{\textbf{US-II} } \\
        \cmidrule(lr){2-3} \cmidrule(lr){4-5}\\
        & DCSI $\uparrow$ & SC $\uparrow$ & DCSI $\uparrow$ & SC $\uparrow$ \\
        \midrule
        DBSCAN & $0.363 \pm 0.004$ & $0.594 \pm 0.006$ & $0.258 \pm 0.001$ &  $0.454 \pm 0.004$  \\
        FairDen  & $0.130 \pm 0.001$ &  $0.494 \pm 0.006$ &  $0.139 \pm 0.001$ &  $0.420 \pm 0.002$   \\
        K-means &  $0.208 \pm 0.002$ & $0.426 \pm 0.005$ & $0.091 \pm 0.000$ & $0.139 \pm 0.001$  \\
        FairSC & $0.153 \pm 0.002$ & $0.089 \pm 0.000$ & $0.087 \pm 0.001$ & $0.263 \pm 0.003$  \\
        \bottomrule
    \end{tabular}
    \label{ustab}
\end{table}
\paragraph{Generative modeling}
\begin{table}[ht]
    \caption{Benchmark of generative models. We evaluated generative models taking into considerations 3 aspects of excelling synthetic samples: quality metric $F_{\alpha, \beta}$, $\alpha$-Precision and $\beta$-Recall. We report mean with std for $7$ runs. }
    \centering
    \small
    \begin{tabular}{lcccccc}
        \toprule
        & \multicolumn{3}{c}{\textbf{AGORA}} & \multicolumn{3}{c}{\textbf{CARVEME}} \\
        \cmidrule(lr){2-4} \cmidrule(lr){5-7}\\
        & $F_{\alpha, \beta}$ $\uparrow$ & $\alpha$-Precision $\uparrow$ & $\beta$-Recall $\uparrow$ & $F_{\alpha, \beta}$ $\uparrow$ & $\alpha$-Precision $\uparrow$ & $\beta$-Recall $\uparrow$ \\
        \midrule
        TabDiff  &   $0.82 \pm 0.00$ & $0.83 \pm 0.00$ & $0.88 \pm 0.00$ &  $0.90 \pm 0.01$ & $0.88 \pm 0.02$ & $0.92 \pm 0.02$ \\
        TabDDPM & $0.70 \pm 0.01$ & $0.67 \pm 0.00$ & $0.87 \pm 0.02$ & $0.67 \pm 0.00$ & $0.54 \pm 0.02$ &  $0.91 \pm 0.01$ \\
        CTGAN & $0.74 \pm 0.00$ & $0.69 \pm 0.00$ & $0.80 \pm 0.00$ & $0.88 \pm 0.00$ & $0.95 \pm 0.00$ & $0.84 \pm 0.01$ \\
        TVAE &  $0.88 \pm 0.00$ &  $0.91 \pm 0.01$ & $0.85 \pm 0.00$ &  $0.89 \pm 0.01$ &$0.94 \pm 0.01$ & $0.83 \pm 0.01$ \\
        \bottomrule
    \end{tabular}
    \label{tab:tree_models}
\end{table}
We tested our approach to generate synthetic competitive environments,
gathered in dataset \texttt{Gen}, by benchmarking current state-of-the-art tabular generative models TabDDPM \cite{kotelnikov2023tabddpm}, TabDiff \cite{shi2025tabdiff}, CTGAN and TVAE \cite{DBLP:journals/corr/abs-1907-00503}. 
We evaluated each model's ability to produce synthetic data resembling the real training samples by calculating quality metrics $\alpha$-Precision, $\beta$-Recall, and $F_{\alpha, \beta}$ as their weighted geometric mean \cite{qian2023synthcity}. 
We also calculated diversity and novelty metrics, following \cite{xiao2022tacklinggenerativelearningtrilemma}, to ensure that the environments we generated are diverse enough to cover the variability of possible competitive environments and novel, i.e., different from the input competitive environments they were trained on (see Supplementary M).

Table \ref{tab:tree_models} shows a comparison of the different generative models.
In general, the results were positive: all models had a quality score $F_{\alpha, \beta}>0.7$ indicating that they were capable of capturing the variability of different environments.
Going forward, we could use this data to assist and build upon the analyses in previous sections, for example, integrating it into the framework of Algorithm \ref{al2}.
\subsection{Limitations and Implications for Future Work}
\label{limitations}
\paragraph{Metabolic models are still in development}
While there are popular tools for assembling metabolic models, the problem of model construction is not ``solved.''
Currently, model construction involves using genomic data to create an initial model and then making modifications in an ad hoc manner to better match model predictions with empirical observations.
This requires that the necessary empirical tests can be conducted on the corresponding organism, which for bacteria is especially rare since the vast majority cannot be grown under laboratory conditions. 
A related limitation is that few models include any kind of gene regulation.
Since gene regulation determines the set of metabolic reactions that are actually used in a given context, the absence of regulatory information means that metabolic models can only highlight what is possible--- not necessarily what is realized.
Given these limitations, it is somewhat surprising how well metabolic models actually predict experiments.
For our dataset, we circumvent these limitations by focusing on the types of questions that metabolic modeling can currently address.
We expect future research into metabolic model assembly to result in more standardized methods of building high-quality, predictive models.
We envision that this development will increase the scope of addressable questions and the value of large-scale metabolic datasets, such as this one.
\paragraph{The biological context is constrained}
When constructing this database, we focused exclusively on interactions between pairs of actively growing bacteria.
In part, this was because such environments could be produced en mass with current computational power and techniques such as flux balance analysis. 
It was also done with an eye towards simplicity.
Real microbial communities span a wide array of types of interactions, including warfare and cooperation \cite{ghoul_ecology_2016,granato2019evolution,d2018ecology}.
Moreover, they often include many interacting species and context-dependent interactions (so-called ``higher order'' interactions) \cite{beyond_pairwise_2017,piccardi_toxicity_2019, mickalide_2017,aguilar_2023,friedman2017community}. 
Including all of this complexity would introduce formidable computational and theoretical challenges, as well as lead to a larger set of arbitrary choices concerning parameters and implementation.
There may be some value in considering only pairs of bacteria, as recent papers have indicated that there might be simple rules that determine the structure of larger communities based on pairwise interactions \cite{garcia2021metabolic, venturelli_deciphering_2018}.
Supporting this, our supervised learning results show that we might be able to identify general across-species features of competitive and cooperative environments, while our transfer learning results suggest these could be generalizable.
\paragraph{What is being learned?} 
An important aspect of any applied machine learning experiment is whether we can interpret what a model learned and whether it applies to the real world.
Our database is intended to be used to address questions concerning the structure of problems faced by real microbes, e.g., can they determine the nature of an interaction based primarily on changes in compound concentrations.
The assumption is that the complexity of metabolism  makes these problems challenging, and there is unlikely to be simple loopholes that learning algorithms can exploit.
While this assumption remains unverified, we conducted primal feature importance of our baseline models to identify the most important compounds from a machine-learning perspective (see Supplementary M). 
There are opportunities here for more advanced techniques, e.g., those inspired by reinforcement learning \cite{rl}, to explore these assumptions.
Even if the assumptions turn out to be true, there is still the question of whether real organisms can actually implement the same kind of learning as found in a machine learning algorithm.
Evaluating this question could lead to the development of new kinds of learning algorithms or insights into the limits faced by biological systems.

\section{Conclusions}
In this paper, we presented Friend or Foe\footnotemark, the largest compendium of datasets curated for machine learning tasks addressing whether bacteria compete or cooperate in different environments.
We benchmarked state-of-the-art machine learning models for four distinct tasks, demonstrating their use in uncovering the eco-evolutionary dynamics of competition and cooperation between bacteria.
Together, these datasets and benchmarks showcase a novel application of machine learning and its potential to uncover fundamental insights into microbial interactions.
\footnotetext{All data (\url{https://huggingface.co/datasets/powidla/Friend-Or-Foe}) and code (\url{https://github.com/powidla/Friend-Or-Foe}) underlying this study are publicly available.}


\par
\begin{ack}


This work was supported by the SciLifeLab \& Wallenberg Data Driven Life Science Program (grant KAW 2020.0239 to LMC and a DDLS Academic PhD grant to EL and LMC).  
The machine learning computations and data handling were enabled by the Berzelius resource provided by the Knut and Alice Wallenberg Foundation at the National Supercomputer Centre.
We also thank the High Performance Computing Center North (HPC2N) at Umeå University for providing computational resources for metabolic modeling.
\end{ack}
\begin{small}
\begingroup
\fontsize{9pt}{12pt}\selectfont
    \bibliography{example_paper}
\bibliographystyle{neurips2025}
\endgroup
\end{small}

\end{document}